\documentclass{aastex}
\usepackage{emulateapj5}
\shorttitle{The SNR G106.3+2.7 and its Pulsar Wind Nebula}
\shortauthors{Kothes et al.}

\begin{document}

%\title{The Comet Supernova Remnant with its ``Boomerang'' Pulsar Nebula}
\title{The SNR G106.3+2.7 and its Pulsar Wind Nebula: \\relics of triggered
star formation in a complex environment}
\author{Roland Kothes\altaffilmark{1}, B\"ulent Uyan{\i}ker\altaffilmark{1},
and Serge Pineault\altaffilmark{2}}
\altaffiltext{1}{National Research Council of Canada, Herzberg Institute of
Astrophysics, Dominion Radio Astrophysical Observatory, P.O. Box 248,
Penticton BC V2A 6K3, Canada}
\email{roland.kothes@nrc.ca, bulent.uyaniker@nrc.ca}
\altaffiltext{2}{D\'epartement de Physique and Observatoire du Mont M\'egantic, 
Universit\'e Laval, Ste-Foy, PQ G1K 7P4, Canada}
\email{pineault@phy.ulaval.ca}

\begin{abstract}
We propose that the pulsar nebula associated with the pulsar
J2229+6114 and the supernova remnant (SNR) G106.3+2.7 are the result
of the same supernova explosion. The whole structure is located at the edge 
of an \ion{H}{1} bubble with extended regions of
molecular gas inside. The radial velocities of both 
the atomic hydrogen and the molecular material suggest a distance of 800~pc. 
At this distance the SNR is 14~pc long and 6~pc wide.
Apparently the bubble was created by the stellar wind and supernova explosions
of a group of stars in its center which also triggered the formation
of the progenitor star of G106.3+2.7. 
The progenitor star exploded at or close to the
current position of the pulsar, which is at one end of the SNR rather
than at its center. The expanding shock wave of the supernova
explosion created a comet shaped supernova remnant by running into dense
material and then breaking out into the inner
part of the \ion{H}{1} bubble. A 
synchrotron nebula with a shell-like structure (the ``Boomerang'') of length 0.8~pc was 
created by the pulsar wind interacting with the dense ambient medium. 
The expanding shock wave created an \ion{H}{1} shell of mass 0.4~M$_\odot$ 
around this nebula by ionizing the atomic hydrogen in its vicinity.

\end{abstract}


\section{Introduction}

The extended radio continuum source G106.3+2.7 was first classified
as a supernova remnant (SNR) by \cite{joncas}. This SNR
was later described by \cite{pineault} in
more detail. They separated the comet shaped structure into a small head
with high radio surface brightness and a large extended tail of low
surface brightness. 
A gap in the atomic hydrogen apparently correlated with the SNR
suggested a distance of about 12~kpc. This would give the SNR a length
of 200~pc and they concluded that G106.3+2.7 is a supernova remnant in the 
late stage of its isothermal evolution.

At the northern edge of the head there is a small shell-like
structure with a diameter of about 3.5'. This source coincides with strong
X-ray emission and the $\gamma-$ray 
source 3EG J2227+6122. It was interpreted as the bow
shock nebula of a fast moving pulsar or a pulsar wind blown bubble
\citep{halpern}. 
These authors estimate the distance as $\approx$ 3~kpc from X-ray
absorption values. Subsequently this pulsar, PSR J2229+6114, was detected
in radio and X-rays \citep{halpern2}. It has a period of 51.6~ms and
a rotational energy loss rate of $\dot{\rm E} = 2.2\cdot 10^{37}$~erg/s.

In this paper, by utilizing the data provided by the Canadian Galactic
Plane Survey (CGPS), we analyze the possible relationship between the pulsar
nebula and the supernova remnant. Since the number of known combined
SNRs is rather low, such an analysis would improve our conceptual
understanding on the interaction of a pulsar nebula with its environment.

\section{Observations and Data Reduction} 

The observations discussed here form part of the CGPS,
which is described in detail by \cite{taylorcgps}. 
In carrying out the research presented here, we have used those 
parts of the CGPS database which derive from observations with the 
DRAO Synthesis Telescope \citep{landecker}
and CO observations from the Five College Radio Astronomy Observatory
(FCRAO) \citep{heyer}. Angular resolution 
varies as
cosec(declination) and therefore changes slowly across the final mosaics.
Before assembly into a mosaic, the data for the
individual fields are carefully processed to remove artifacts and to
obtain the highest dynamic range, using the routines described by
\cite{willis}.
Accurate representation of all structures to the largest scales is
assured by incorporating data from large single antennas with data
from the Synthesis Telescope, after suitable filtering in the Fourier
domain. Continuum data are derived from the 408~MHz all-sky survey of
\cite{haslam}, and from the 1.4 GHz Effelsberg survey \citep{reich}. 
Single antenna \ion{H}{1} data are obtained from a 
survey of the CGPS region made with the DRAO 26-m Telescope 
\citep{higgs}. To improve the resolution of the 1420 MHz 
continuum observations
we also included data taken from the NVSS \citep{nvss}. The NVSS
map was convolved to the resolution of our CGPS data. The resulting map
was subtracted from the CGPS map and the difference added to the original
NVSS map. The final image has a resolution of 45''.

\section{The structure of G106.3+2.7}

In Fig. \ref{radio} the radio image of G106.3+2.7 at 1420~MHz is shown. 
In the lower image the overall 
structure of the source is clearly visible. The major parts
are a comet shaped structure with a diffuse extended tail of low
radio surface brightness and
a smaller more structured head with higher radio surface brightness.
The upper image is an enlargement of the latter. The brightest
feature is the ``Boomerang'' to the north, described by
\cite{halpern}. The point-like source just to its south is most likely
an extragalactic source \citep{halpern}. The head itself shows two prominent
features. There is a peak in the center, south of the extragalactic
source and a shell-like structure on its southern boundary. Also visible
is a diffuse bridge extending from the head to the
pulsar nebula, suggesting an interaction of those structures. The
SNR's radio continuum emission has a spectral index of
$\alpha \approx -0.57$ (S$\sim \nu^\alpha$) \citep{pineault}.
Neither the tail nor the head show significant spectral
fluctuations. However, the pulsar nebula seems to have a flat 
radio spectrum \citep{halpern}. We obtained a flux density
of $90 \pm 5$~mJy by integration of our 1420~MHz data. Together with a flux density
of 80~mJy at 4850~MHz taken from \cite{becker} we derive a spectral index
of $\alpha = -0.1 \pm 0.1$ for the pulsar nebula. This confirms
the flat radio spectrum proposed by \cite{halpern} and supports its
interpretation as a pulsar-powered synchrotron nebula.

Fig. \ref{polar} displays a map of polarized intensity at 1420~MHz
in greyscale together with white contours representing
the total power radio continuum. We find linear polarization
in all parts of the object. The tail has only a small
number of polarized patches but these show high percentage
polarization of up to 70~\%. The head is polarized
except for an area between the pulsar nebula and the center
of the head where we also find a peak in total power emission.
The head shows peak polarization of up to 70~\%. The most
prominent polarization feature, however,  is the pulsar nebula with an integrated
percentage polarization of 40~\%. Parts of the
radio continuum bridge to the pulsar nebula appear to be highly polarized.
These high percentage polarizations indicate low depolarization
between the SNR and us. The mean rotation measure towards the ``Boomerang''
and the SNR G106.3+2.7 is around $RM \approx -200$~rad/m$^2$. This high
rotation measure results in Faraday rotation of $\approx 500^\circ$
at a frequency of 1420~MHz. Since we find only low depolarization on
the ``Boomerang'' it requires a very small distance which 
makes this object a local phenomenon, probably within 2~kpc.
If it was further away, beam depolarization would occur because of
superimposed regions of differing Faraday rotation within the beam.

The large OB association Cep OB2, which is located at a distance
of about 950~pc \citep{garmany}, lies in the direction of G106.3+2.7
and partly overlaps with the supernova remnant. The members of this
OB association ionize the material between them which leads to
the depolarization of extended polarized emission from behind them
at low radio frequencies. This gives us another upper limit for
the distance to G106.3+2.7.
\section{The structure of the cold environment}

\subsection{The head's hydrogen envelope}

In Fig. \ref{hichan} we plot channel maps of our \ion{H}{1} data cube
representing an area around the head of G106.3+2.7 at low negative
radial velocities. There is a shell-like \ion{H}{1} structure around the
eastern edge of the head best seen at -6.41~km/s but also
quite prominent at -5.59~km/s and -7.23~km/s. A shell of \ion{H}{1}
is wrapped around the ``Boomerang'', peaking at the same velocity
as the shell around the head of the SNR. The coincidence in
velocity clearly demonstrates that the SNR and the pulsar nebula
are associated. This is the first time that an interaction
between a pulsar nebula and cold interstellar material has been observed.
It looks like the pulsar is either pushing the neutral hydrogen outwards
or ionizing cold atomic hydrogen in
its vicinity with its pulsar wind and thus creating the shell-like structure. 

The center velocity
of -6.4~km/s puts the SNR at a distance of 800~pc. At this
distance the SNR has a length of 14~pc and a maximum width of 6~pc.
The pulsar nebula is 0.8~pc wide. The small shell of atomic
hydrogen contains about 0.4~M$_\odot$ and the bigger shell around the head
about 5~M$_\odot$. This leads to densities of about 200~cm$^{-3}$ within
both structures. 

\cite{halpern} calculated the foreground \ion{H}{1} column density
to be $N_H = 6.3\cdot 10^{21}$~cm$^{-2}$ from the spectral fit to
their X-ray data. Based on this value they estimated a distance of
3~kpc. We integrated the foreground \ion{H}{1} column density from our
data using $N_H [{\rm cm}^{-2}] = 1.823\cdot 10^{18} \int T_B(v) dv$ 
\citep{kerr} with $T_B$ representing the \ion{H}{1} excess brightness
temperature and $v$ its radial velocity. The result is 
$N_H \approx 2\cdot 10^{21}$~cm$^{-2}$. This
equation, however, is only valid if the neutral hydrogen in the 
foreground is optically thin. Comparison of the brightness
temperature of two strong point like background sources in the vicinity
of G106.3+2.7 ($l=105.68^\circ$, $b=2.42^\circ$ and $l=107.84^\circ$,
$b=2.31^\circ$) with their
\ion{H}{1} absorption profiles indicates that most of the foreground
material is optically thick with opacities higher than 3. This easily
explains the factor of 3 between the calculated foreground $N_H$ and
the one derived from the X-ray data.

\subsection{The head's molecular envelope}

In Fig. \ref{cochan} we plot channel maps of our CO data cube
in the same area and at the same radial velocities as the \ion{H}{1} data
in Fig. \ref{hichan}. At the same velocities where we found the eastern
hydrogen envelope there is a molecular envelope around the western edge
of the head structure. But in this case we do not see the same structure
in all channels. The location of the CO envelope is to the west at -5.6~km/s
but rotates to the northwest at -7.2~km/s. Since around the eastern edge we find neutral
hydrogen in all channels we can assume that the hydrogen completely
surrounds this area. The CO, on the other hand, seems to be incomplete
giving the expanding supernova remnant space to break through and form
the smooth extended tail. The narrow shell of molecular material
contains about 40~M$_\odot$ of molecular hydrogen assuming
$N_{H_2} [{\rm cm}^{-2}] = 2.3\cdot 10^{20}\cdot T_{CO}\cdot \delta v$.
Here $N_{H_2}$ is the column density of molecular hydrogen, $T_{CO}$ [K]
is the CO brightness temperature and $\delta v$ [km/s] is the velocity
width of these structures. 

\subsection{The big picture}

Fig. \ref{bigbubble} shows an image of the entire region surrounding
G106.3+2.7. The head of the SNR and its molecular and atomic envelope
are visible in the lower left part of the image as is the small shell
surrounding the pulsar wind. Apparently these structures represent only
a small part of the entire cold environment. There is a large bubble
centered at $l\sim 106.1^\circ$ and $b\sim 3.1^\circ$, 
with an inner diameter of about $1^\circ$ which translates to about 14~pc
at a distance of 800~pc. The outer diameter is about 20~pc. 
At the northern and eastern edge of the bubble
we find patches of molecular material. The supernova remnant has
apparently created a little bubble at the lower left edge
of the large bubble with its shock wave.

\section{Discussion}

\subsection{The creation of the cold environment}

We suggest that a first generation of stars was formed in the
center of the big \ion{H}{1} bubble which at that time was a cloudy 
dense medium. With stellar wind effects and/or supernova explosions 
the large bubble was formed. However, some very dense cloudlets remained
because they were too dense and probably also too far away from the
location of the exciting stars. The fact that the remaining patches
we observe are located at the outer edges of the bubble supports this,
as does an \ion{H}{1} structure surrounding the northeastern
patch of molecular material, which is most likely the remains of 
dissociated molecular hydrogen at the surface of this cloud. 
Stellar winds and supernova explosions within the large bubble
apparently also triggered the
creation of the progenitor star of our SNR G106.3+2.7.
At the time the progenitor star of G106.3+2.7 was formed, the environment
most likely looked very much like the two northern patches of
molecular material. The new born star was located inside this 
molecular environment. It was a massive star, probably an early
B or an O-type star. The small bubble in the cold environment
in which we now find the head of the SNR was either formed by the
progenitor star's stellar wind or by the expanding shockwave of the 
supernova explosion itself.

\subsection{A bow shock or a pulsar wind nebula}

\def\binf{B_{\infty}}

Usually pulsar nebulae are expanding inside a bubble created by the shock wave
of a supernova explosion. In this case there is no material
left to interact with, because all the matter has been swept up and carried
away already. The pulsar in G106.3+2.7 is apparently interacting
with its dense cold environment. In the following we investigate
two possible scenarios leading to the shell-like shape of
the ``Boomerang'' pulsar nebula. 

\subsubsection{The ``Boomerang'' as a bow shock nebula}

Since the shape of the object bears a striking resemblance to
a bow-shock structure, we first investigate the constraints
which are imposed by the observations in order for this hypothesis to
be viable.  For simplicity, we assume
that the explosion which caused the SNR took place near the
center of the head component and resulted in the candidate pulsar
getting a kick velocity which took it to its present position
near the bow structure.  In view of the observed morphology,
we also suppose the motion to be nearly in the plane of the sky
and neglect projection effects. The distance travelled, $L$, is 
comparable to the radius of the head component $R_s$, so we set
$L \approx R_s$.  Letting $t$ represent the pulsar travel time
and SNR age in years, $v_p$ the pulsar space velocity in km/s, and
$R_s$ in pc,  we have 

\begin{equation}
R_s \approx 1.0\cdot 10^{-6} v_p t
\end{equation}

The equality between the pulsar relativistic wind
pressure and ISM ram pressure implies that
\begin{equation}
\frac{{\dot E}}{4 \pi l^2 c} = n_o \mu\, m_H v_p^2 \Rightarrow 
\dot{E}_{36} \approx 6.2 \mu n_o l^2 v_p
\end{equation}
Here
$n_o$ is the ambient number density in cm$^{-3}$, $l$ is the distance between 
the pulsar and the apex of the bow shock in pc, and ${\dot E}_{36}$ is the pulsar
rotational energy loss rate in $10^{36}$~erg/s. SNR dynamics give us:
\begin{equation}
R_s= \left\{ \begin{array}{ll}
0.34 \left (\frac{E_{51}}{\mu n_o}\right )^{0.2} t^{0.4} & \hbox{for adiabatic expansion}\\[0.2cm]
1.4 \left (\frac{\epsilon E_{51}}{\mu n_o}\right )^{5/21} t^{2/7} & \hbox{for isothermal expansion}\\
\end{array}\right. 
\end{equation}
Here the parameter $\epsilon$ is the ratio of thermal energy in the SNR
to the initial explosion energy $E_{51}$ [$10^{51}$~erg].

The pulsar distance from the center of the head is about 1.5~pc, its distance from
the apex of the bowshock is about 0.4~pc and its rotational energy loss rate
${\dot E} = 2.2\cdot 10^{37}$~erg/s \citep{halpern2}. With an estimated value of 0.3
for $\epsilon$ we get:

\begin{equation}
E_{51} \approx \left\{ \begin{array}{ll}
1.7\cdot 10^{-8} v_p & \hbox{adiabatic case}\\[0.2cm]
1.2\cdot 10^{-6} v_p^{0.2} & \hbox{isothermal case}
\end{array}\right. 
\end{equation}

Any reasonable value
for the pulsar velocity (100~km/s $\le v_p \le$ 1000~km/s) will give us  
by far the lowest ever recorded explosion energy for the supernova
explosion. Allowing for inclination effects or substantial 
variations in the observed parameters still leaves us with 
highly unlikely values of the energy parameters. 

\subsubsection{The ``Boomerang'' as a pulsar wind nebula}

As an alternative, \cite{halpern} have suggested that the bow-shaped structure
can be confined by the thermal pressure of the surrounding medium.
This is indeed possible, but it implies an anomalously low
velocity for an object belonging to a high-velocity class.
The fact that the spectral index is flat makes it almost unavoidable
that the emission originates from the relativistic pulsar wind 
itself.  However we believe that the confinement is simply achieved
because the relativistic wind is encountering the dense SNR shell in
the northeast.  In our picture the SN explosion took place at or very
near the present position of the pulsar.  In the northeast the shock
has been quickly decelerated by its encounter with a 
relatively nearby and particularly high density medium, 
whereas towards the south it has expanded into a moderately dense
medium (SNR head component) while breaking out
in the southwest into a region of much
lower density and giving rise to the tail component. 

The morphology
of the head and the tail structure supports this hypothesis. The head 
shows only the ``Boomerang'' and the southern shell as indications
for the encounter of a shockwave with dense material. We do not find
a shell structure towards the east of the head but we find an \ion{H}{1}
shell all around the east and the south of the head. We can explain this
by moving the location of the explosion to the current position of the 
pulsar. To the north and to the east the shockwave is running into dense
material. The first free path for the shockwave is to the southeast where 
the southern
shell begins. Another point for this scenario is that the distance between
the location of the supernova and the shell to the south is now larger
that its distance to the beginning of the tail, which should be expected
in our picture. If the explosion occured in the center of the head this
would not be the case.

What seems peculiar however is the fact that the little 
bow-shaped shell in the north does not show enhanced non-thermal
radio emission. This would appear to go against the expected
view that an SNR should be brighter where the blast wave is
moving against dense regions of the ISM.  A similar situation has
been encountered by \cite{pineault2} in their analysis of the
CTA1 SNR.  These authors argued that, in overdense regions where
the shock velocity is considerably reduced, the
Fermi acceleration mechanism \citep{bell} might either be too slow or 
simply break down.  The time efficiency can be quantified 
by evaluating the timescale
$\tau_{\rm acc}$ for acceleration of an electron to energy $E$, given
by $\tau_{\rm acc} \propto E/B\,v^2$ \citep{ellison,reynolds}.
Hence either a
low shock velocity or low ambient magnetic field, or a combination of
both may imply that the electrons radiating at a given observing
frequency $\nu$, where $E \propto (\nu/B)^{1/2}$,
are not yet present in the energy distribution.
An alternative approach is to consider the growth and decay rates
of the Alfv\'en waves responsible for the repeated scattering of the
relativistic particles back and forth across the shock \citep{bell,blandford}.
On one hand, the growth rate is proportional to the shock velocity
and thus decreases as the shock slows down and, on the other hand,
the decay rate depends on the first power of the number density
of neutral atoms.  Again this means that the acceleration
mechanism may be quenched whenever the shock wave encounters 
sufficiently dense clouds.
It is also possible that a thin steep (spectral index of order -0.5)
non-thermal layer is present but not detectable because of
beam smearing.

The interpretation of the tail component as resulting from a
breakout of the SNR shock wave is also compatible with theoretical
predictions.  The numerical studies of breakout by \cite{tenorio}
have shown that the asymptotic ratio of the velocity of the material 
breaking out into a region of lower density to that of
the shock wave still propagating into the original medium could
be expressed as 
\begin{equation}
\binf = 4.22 - 1.89~(\log_{10} \lambda +2),
\end{equation}
where $\lambda \le 1$ is the ratio of the densities.  The same ratio
at any normalized time $\tau = t/t_b$, where $t_b$ is the time at
which breakout occurred, was shown by \cite{pineault3}
to be satisfactorily represented by the
approximate analytical relation 
\begin{equation}
B(\tau) = \binf - (\binf -1)/\tau.
\end{equation}
It follows from this relation that the ratio $ r = L_t/R_s \approx \binf$
where $L_t$ is the length of the tail component, as long as $\tau >> 1$.
Using $R_s\approx 3$~pc and $L_t\approx 12$~pc gives us a value of 4
for $r$. The length of the tail is difficult to determine since there
is no sharp outer boundary.
The observed value implies $\lambda
\approx 0.01$.  A value of 8 for $r$ would correspond
to $\lambda \approx 10^{-4}$.  It can also be shown that the breaking
out material attains a maximum velocity 
\begin{equation}
v_{bm} = v_b \, [0.35 \, \binf \, (\binf/(\binf -1))^{0.6}]
\end{equation}
 at $\tau_m = (8/3) \, (1 - 1/\binf)$,
where $v_b$ is the velocity just before breakout.  For $\binf = 4$ , 
the maximum velocity is about $1.7\,v_b$. 
Assuming that the age of the SNR is comparable to the characteristic age
of the pulsar we take $t\approx 10460$~yr \citep{halpern2}. For an adiabatic
expanding SNR we derive $E_{51}/{\mu n_0} \approx 4.9\cdot 10^{-4}$. 
With a density of $n_0\approx 100$~cm$^{-3}$ the explosion energy would be
$E \approx 7\cdot 10^{49}$~erg. Usually the characteristic age of a pulsar is higher
than its real age which means the calculated explosion energy is a lower limit.
From the length of the tail, an average velocity 
of $1125 \rm\, km\,s^{-1}$ is inferred.  
Equating this to the maximum velocity
taken as $1.7 \, v_b$ gives $t_b \approx 540$~yr, corresponding to 
$R_{sb} \approx 1$~pc.  Although these numbers should not be taken at face
value, they nevertheless indicate that the breakout hypothesis is
consistent with the observed morphology.

\section{Conclusion}

Using  the  data from  the  CGPS we  studied  the  morphology and  the
environment  of the SNR  G106.3+2.7 and  the pulsar nebula associated
with  the pulsar J2229+6114 (the  ``Boomerang''). We
conclude that these  sources are the result of the  same SN event.
The SNR  morphologically consists of  two parts giving the  object its
cometary  shape, in  which the  tail is  due to  an outbreak  into the
interior of  a big \ion{H}{1} bubble,  whereas the head  is created by
the expanding shock wave interacting with dense ambient material. 
The  kinematics  of  the  associated neutral  hydrogen  and  molecular
material,  $v_{kin}\sim -6.4$~km/s,  suggests that  both objects  are
located at the same distance of 800~pc, thereby
the  SNR is 14~pc long  and 6~pc wide  and the  pulsar nebula  has a
diameter of  0.8~pc. This close distance is also
supported by the presence of high polarized emission.

By  studying the energetics  of the SNR  we found
that the ``Boomerang''  nebula is created by the  relativistic wind of the
pulsar. For the ``Boomerang''  we found  a spectral  index of  $\alpha =
-0.1$ and $\sim$ 40~\%  integrated percentage polarization at 1420~MHz.
The analysis of the cold environment suggests that  the creation of the
progenitor star was triggered by  stellar wind and/or SN explosions of
a group of stars which also created the \ion{H}{1} bubble. 
In this way this is the first detection of a pulsar nebula apparently
interacting with its cold environment. It is also an example of a pulsar
``displaced'' from the center of its associated SNR. In this case,
however, the displaced position of the pulsar is not due to high 
pulsar velocity.

\acknowledgments{
The Dominion Radio Astrophysical Observatory is a National Facility
operated by the National Research Council.  The Canadian Galactic
Plane Survey is a Canadian project with international partners, and is
supported by the Natural Sciences and Engineering Research Council
(NSERC). SP was supported by NSERC and the Fonds FCAR of Qu\'ebec.
We thank Lloyd Higgs and Tom Landecker for careful reading of the
manuscript and discussions.}

\clearpage

\begin{figure*}
   \caption{The SNR G106.3+2.7 at 1420 MHz radio continuum. The \ion{H}{2} region
   S142 is indicated as are the main part of the SNR and the pulsar wind
   nebula. Contours are at 6.6, 6.8, 7.0, 7.2, 7.5, 7.8, 8.1, 8.4, and 8.7~K
   in the lower map and from 7~K to 11~K in steps of 0.5~K in the upper
   image. The resolution is 45''.}
   \label{radio}
\end{figure*}

\begin{figure}
   \caption{Greyscale plot of the polarized emission of the SNR G106.3+2.7 
    at 1420 MHz. White contours represent the total power radio continuum
    emission at 1420 MHz. Contour levels are 100, 200, 500, 800, and 1200~mK. 
    Both maps were convolved to a resolution of 2' to 
    increase the signal to noise ratio in the polarized intensity map.
    Contours for the polarized intensity (black) are from 100~mK to 500~mK in steps
    of 50~mK.}
   \label{polar}
\end{figure}

\begin{figure}
   \caption{Greyscale plots of \ion{H}{1} channel maps at low negative velocities
      with velocities indicated. White contours 
      represent total power 
      radio continuum at 1420 MHz. Both maps were convolved to a resolution 
      of 2' to increase the signal to noise ratio in the \ion{H}{1} maps.}
   \label{hichan}
\end{figure}

\begin{figure}
   \caption{Greyscale plots of CO channel maps at low negative velocities
      with velocities indicated. White contours, as in Fig. \ref{polar}, represent total power 
      radio continuum at 1420 MHz. Both maps were convolved to a resolution 
      of 2' to increase the signal to noise ratio in the CO maps.}
   \label{cochan}
\end{figure}

\begin{figure}
   \caption{Greyscale plot of neutral hydrogen associated with the
      SNR G106.3+2.7. Overlaid black contours represent molecular material
      and the white contours (at 200, 500, 800, and 1200~mK) radio continuum at 1420~MHz.
      All data have been convolved to a resolution of 2' to improve the signal 
      to noise ratio. For the neutral hydrogen and the CO the three
      channels at -5.6~km/s, -6.4~km/s, and -7.2~km/s were averaged
      together.}
   \label{bigbubble}
\end{figure}


\begin{thebibliography}{}
   
   \bibitem[Bell (1978)]{bell} Bell A.R., 1978, MNRAS 182, 147
   
   \bibitem[Becker et al.(1991)]{becker} Becker R.H., White R.L., Edwards A.L.,
      1991, ApJS 75, 1
   
   \bibitem[Blandford \& Ostriker(1978)]{blandford} Blandford R., 
      Ostriker J.P., 1978, ApJ 221, L29
   
   \bibitem[Condon et al.(1998)]{nvss} Condon J.J., Cotton W.D., Greisen E.W.,
      Yin Q.F., Perley R.A., Taylor G.B., Broderick J.J., 1998, AJ 115, 1693

   \bibitem[Ellison et al.(1990)]{ellison} Ellison D.C., Jones F.C.,
      Reynolds S.P., 1990, ApJ 360, 702
   
   \bibitem[Garmany \& Stencel(1992)]{garmany} Garmany C.D., Stencel R.E., 1992,
      A\&AS 94, 211
   
   \bibitem[Halpern et al.(2001a)]{halpern} Halpern J.P., Gotthelf K.M., Leighly K.M., 
      Helfand D.J., 2001, ApJ 547, 323

   \bibitem[Halpern et al.(2001b)]{halpern2} Halpern J.P., Camilo F., Gotthelf E.V., 
      Helfand D.J., Kramer M., Lyne A.G., Leighly K.M., Eracleous M., 
      2001, ApJ 552, L125

   \bibitem[Haslam et al.(1982)]{haslam} Haslam C.G.T., Stoffel H., Salter C.J., Wilson W.E.,
      1982, A\&AS 47, 1
   
   \bibitem[Heyer et al.(1998)]{heyer} Heyer M.H., Brunt C., Snell R.L., 
      Howe J.E., Schloerb F.P., Carpenter J.M., 1998, ApJS 115, 241
   
   \bibitem[Higgs \& Tapping(2000)]{higgs} Higgs L.A., Tapping K.F., 2000, AJ 120, 2471

   \bibitem[Joncas \& Higgs(1990)]{joncas} Joncas G., Higgs L.A., 1990, A\&AS 82, 113

   \bibitem[Kerr(1968)]{kerr} Kerr F.J., 1968, In: Nebulae and Interstellar 
      Matter, Stars and Stellar Systems Vol. III, p575, B.M. Middlekorst and 
      L.H. Aller (eds)
   
   \bibitem[Landecker et al.(2000)]{landecker} Landecker T.L., Dewdney P.E., Burgess T.A.,
      Gray A.D., Higgs L.A., Hoffmann A.P., Hovey G.J., Karpa D.R.,
      Lacey J.D., Prowse N., Purton C.R., Roger R.S., Willis A.G.,
      Wyslouzil W., Routledge D., Vaneldik J.F., 2000, A\&AS 145, 509

   \bibitem[Pineault et al.(1987)]{pineault3} Pineault S., Landecker T.L.,
      Routledge D., 1987, ApJ 315, 580
   
   \bibitem[Pineault et al.(1997)]{pineault2} Pineault S., Landecker T.L.,
      Swerdlyk C.M., Reich W., 1997, A\&A 324, 1152

   \bibitem[Pineault \& Joncas(2000)]{pineault} Pineault S., Joncas G., 2000, AJ 120, 3218
   
   \bibitem[Reich et al.(1997)]{reich} Reich W., Reich P., F\"urst E., 1997, A\&AS
      126,413

    \bibitem[Reynolds(1996)]{reynolds} Reynolds S.P., 1996, ApJ 459, L13
    
    \bibitem[Taylor et al.(2001)]{taylorcgps} Taylor A.R., Dewdney P.E., Landecker T.L.,
      Martin P.G., Brunt C., Dougherty S.M., Durand D., Gibson S.J.,
      Gray A.D., Higgs L.A., Kerton C.R., Knee L.B.G., Kothes R., 
      Peracaula M., Purton C.R., Uyan{\i}ker B., Wallace B.J., 
      Willis A.G., 2001, AJ, submitted
 
   \bibitem[Tenorio-Tagle et al.(1985)]{tenorio} Tenorio-Tagle G., 
      Bodenheimer P., Yorke H.W., 1985, A\&A 145, 70
   
   \bibitem[Willis(1999)]{willis} Willis A.G., 1999, A\&AS 136, 603

\end{thebibliography}
\end{document}